\documentclass[twoside,a4paper,11pt]{sea10}
\usepackage{graphicx}
\usepackage{hyperref}
\begin{document}
\pagenumbering{arabic}
\pagestyle{myheadings}
\thispagestyle{empty}
{\flushleft\includegraphics[width=\textwidth,bb=58 650 590 680]{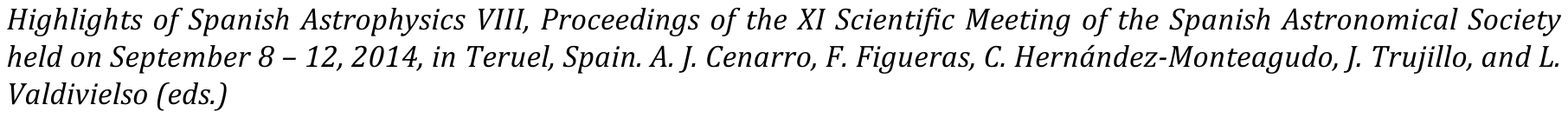}}
\vspace*{0.2cm}
\begin{flushleft}
{\bf {\LARGE
%
Traces of co-evolution in high z X-ray selected and submm-luminous QSOs. 
%
}\\
\vspace*{1cm}
%
A. Khan-Ali$^{1}$,
F.J. Carrera$^{1}$,
M.J. Page$^{2}$,
J.A. Stevens$^{3}$,
S. Mateos$^{1}$,
M. Symeonidis$^{2,4}$
and 
J.M. Cao Orjales$^{3}$
%
}\\
\vspace*{0.5cm}
%
$^{1}$
Instituto de F\'\i{}sica de Cantabria (CSIC-UC), Santander 39005, Spain.\\
$^{2}$
Mullard Space Science Laboratory, University College London, RH5 6NT, UK.\\
$^{3}$
Centre for Astrophysics Research, University of Hertfordshire, AL10 9AB, UK.\\
$^{4}$
University of Sussex, Department of Physics and Astronomy, BN1 9QH, Sussex, UK.
%
\end{flushleft}
%
\markboth{
High X-ray selected and submm-luminous QSOs
}{ 
%
Khan-Ali et al.
%
}
\thispagestyle{empty}
\vspace*{0.4cm}
\begin{minipage}[l]{0.09\textwidth}
\ 
\end{minipage}
\begin{minipage}[r]{0.9\textwidth}
\vspace{1cm}
\section*{Abstract}{\small
%
We present a detailed study of a X -ray selected sample of 5 submillimeter bright QSOs
at $z\sim2$ , where the highest rates of star formation (SF) and further growth of black 
holes (BH) occur. Therefore, this sample is 
a great laboratory to investigate the co-evolution of star formation and AGN. We present here 
the analysis of the spectral energy distributions (SED) of the 5 QSOS, including new data from 
Herschel PACS and SPIRE. Both AGN components (direct and reprocessed) and like Star Formation (SF) 
are needed to model its SED. From the SED and their UV-optical spectra we have estimated the mass 
of the black hole ($M_{BH} = 10^9 - 10^{10} M_{SUN}$) and bolometric luminosities of AGN ($L_{BOL} = (0.8-20) \times 10^{13} L_{SUN}$). 
These objects show very high luminosities in the far infrared range (at the H/ULIRG levels) and very high rates 
of SF (SFR = 400-1400 $M_{SUN}$/y). Known their current SFR and their BH masses, we deduce that 
their host galaxies must be already very massive, or would not have time to get to the local 
relation between BH mass and bulge. Finally , we found evidence of a possible correlation between 
the column density of ionized gas detected in X-rays ($NH_{ion}$) and SFR, which would provide a 
link between AGN and SF processes.

%
\normalsize}
\end{minipage}
%
%
%
\section{Introduction \label{intro}}
In the last two decades, it has become clear that most local
spheroidal galaxy components (elliptical galaxies and the bulges of
spiral galaxies) contain a super massive black hole (SMBH) in their
centres. The proportionality between black hole (BH) and spheroid mass
suggests a direct link between the growth of the black hole as an
Active Galactic Nucleus (AGN) and the stellar mass of the spheroid
(e.g. \cite{marconi04}, \cite{kormendy}). Identifying the main
mechanisms for formation and evolution of galaxies, and their
interrelation to that of the growth of their central black holes is a
major issue in Astrophysics and Cosmology. In this work, we have studied a sample of X-ray-obscured QSOs,
described by \cite{page04} and \cite{stevens05}, and studied by
\cite{stevens04,stevens10}, \cite{page11} and \cite{carrera11}, at
z$\sim$1-3 when most of the SF and BH growth are ocurring in the
Universe. These QSOs have strong submm emission, much higher than typically found in QSOs at similar 
redshifts and luminosities. \cite{stevens05} found detections at $> 5\sigma$ significance at 850$\mu$m (SCUBA). Here, we endeavour to get the physical properties of the central QSOs
(luminosities, BH masses, Eddington ratios, etc.) and their host
galaxies (SFR, $M_{DUST}$, $M_{GAS}$, etc.) and their mutual
relationships (or lack thereof). In addition, we will try to fathom
their place in AGN-host galaxy co-evolution models.

\section{Results}
\label{resultSection}
We have constructed the spectral energy distributions (SED) of all
our objects, in $\nu L_{\nu}$. 
All objects clearly show at least two components: an UV-NIR contribution
attributable to direct accretion disk emission (as expected from their
type 1 nature), intrinsically absorbed in the cases of RX~J0941 and
RX~J1218; and a reprocesed thermal component in the MIR region from
warm optically thick dust further away from the nucleus (the torus).
In all of them we also observe an additional FIR/submm component
associated to cooler dust, heated by star formation (SF). We thus
confirm the presence of strong FIR emission due to SF in these
objects, at the ULIRG/HLIRG level (compared to e.g. Mrk 231).

Our next goal is to make fits to our data with different templates to
extract quantitative information about our objects. 
We have selected
the \textit{Sherpa: CIAO's modeling} \& \textit{fitting package}
\cite{sherpa} module for the Python platform to
perfom the fitting.  SEDs were fitted in $\nu L_{\nu}$ versus
rest-frame $\lambda$.
We use relatively simple empirical and theoretical templates, 
aiming at reproducing the general shape of the SEDs of our objects. 
We do not attempt to extract detailed physical information about our objects 
from those templates, since our data do not warrant such undertaking, and 
the underlying physics is likely to be more complex 
than that considered in the models. We have modelled the distinct constituents  with
three different components:

{\bf  An AGN accretion disk component:} This template models the direct emission 
from the accretion disk of the AGN (sub-parsec scales). We use the pure disk newAGN4 template from \cite{rowan}, 
  affected by intrinsic extinction (see below). The only
  free parameter is the normalization of the template. We have
  normalized the template to its integral in the 0.12-100~$\mu$m range
  (standard limits used to estimate the accretion disk luminosity).
  From the fits we directly obtain the value of the disk luminosity in that range,
  $L_{DISK}$.

{\bf A torus component:} This template models the re-emission from the warm and hot dust 
 (on tens of parsecs scales, beyond the sublimation radius, see \cite{antonucci}) that is warmed by the accretion disk emission. We have used both an empirical
  template from \cite{rowan} (dusttor, based on an average quasar spectrum) and three dusty clumpy torus models
  from \cite{nenkova} found by \cite{roseboom} to represent the
  average properties of type 1 QSOs (torus1, torus2 and torus3). 
  Similarly, we have normalized them to their integral in the
  1-300~$\mu$m range (again, standard limits used for the torus
  luminosity), so the only free parameter is $L_{TORUS}$.

{\bf A SF component:} we have used a subset of the
  \cite{siebenmorgenSB} spherical smooth models, found by \cite{myrto}
  to encompass the observed SEDs of star-forming galaxies at least up
  to $z\sim 2$. The full \cite{siebenmorgenSB} models have five free
  parameters to obtain their 7000 templates. The subset of templates
  recommended by \cite{myrto} (around 2000 templates) are divided into
  11 sub-grids, depending on the maximum radius and different star
  populations. For each source, we found the best-fit template for
  each of these 11 sub-grids.  We did not attempt to extract detailed
  physical information from the particular best-fit templates, since
  instead we were looking for physically-motivated templates that
  reproduced the spectral shape of the data. Therefore, the only
  parameter that we obtained from these fits, was the integrated FIR luminosity (40-500$\mu$m)
  $L_{FIR}$.

For each source (see below) we
have chosen the family of models with the lowest $\chi^2$.
Table~\ref{LumResults} shows the average values of the luminosities
and extinction values among each best-fit family for each source and the minimum
value of $\chi^2$ for each best-fit family. We have estimated the uncertainties in those values from the standard
deviation. We believe that these values and their uncertainties are
fair estimates of the luminosities of each component and of the
effects of the photometric errors and our lack of an accurate physical
model for what is really happening in each source.

\begin{figure*}
  \centering
   \includegraphics[width=0.45\textwidth]{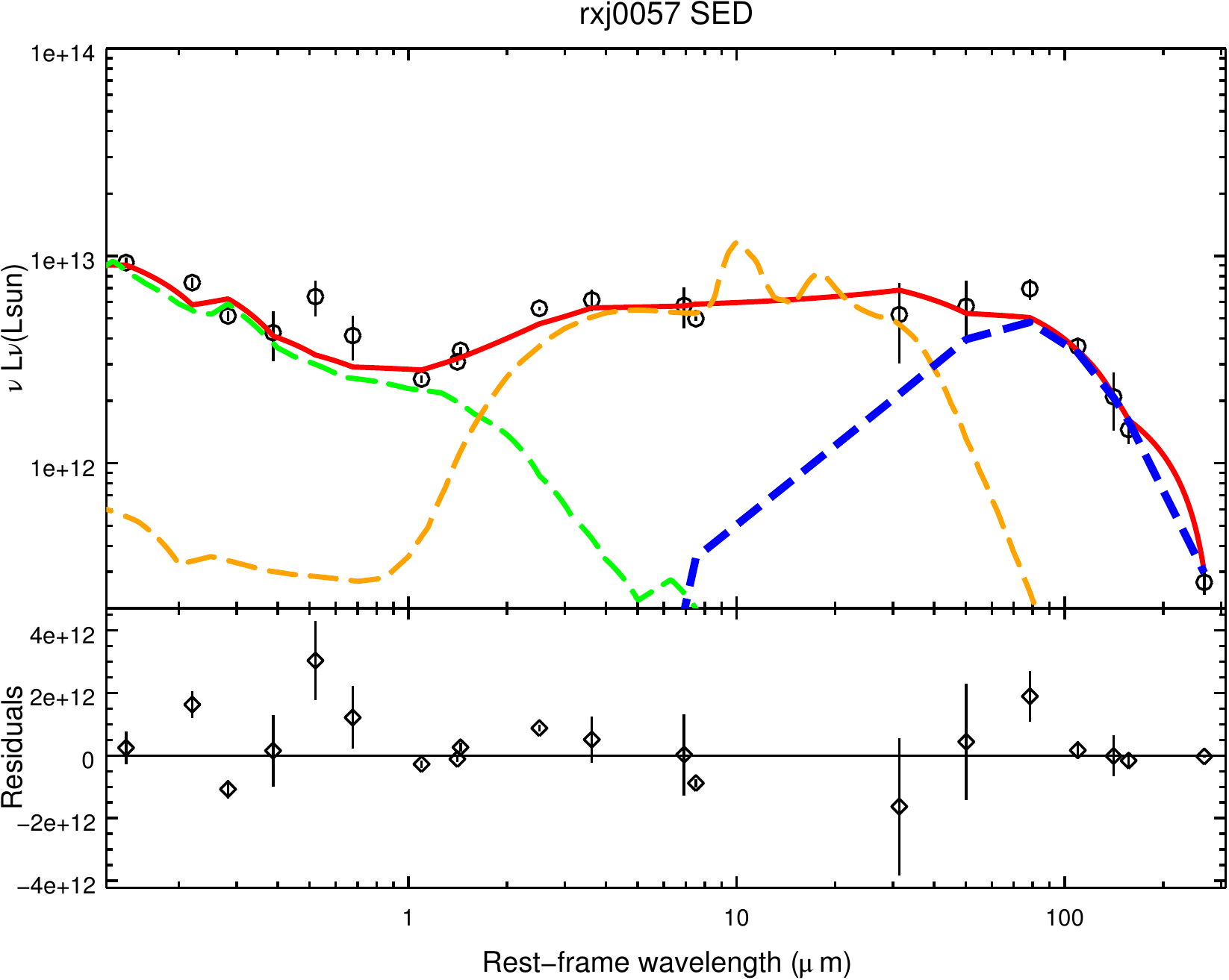} {\hspace {1cm}}
   \includegraphics[width=0.45\textwidth]{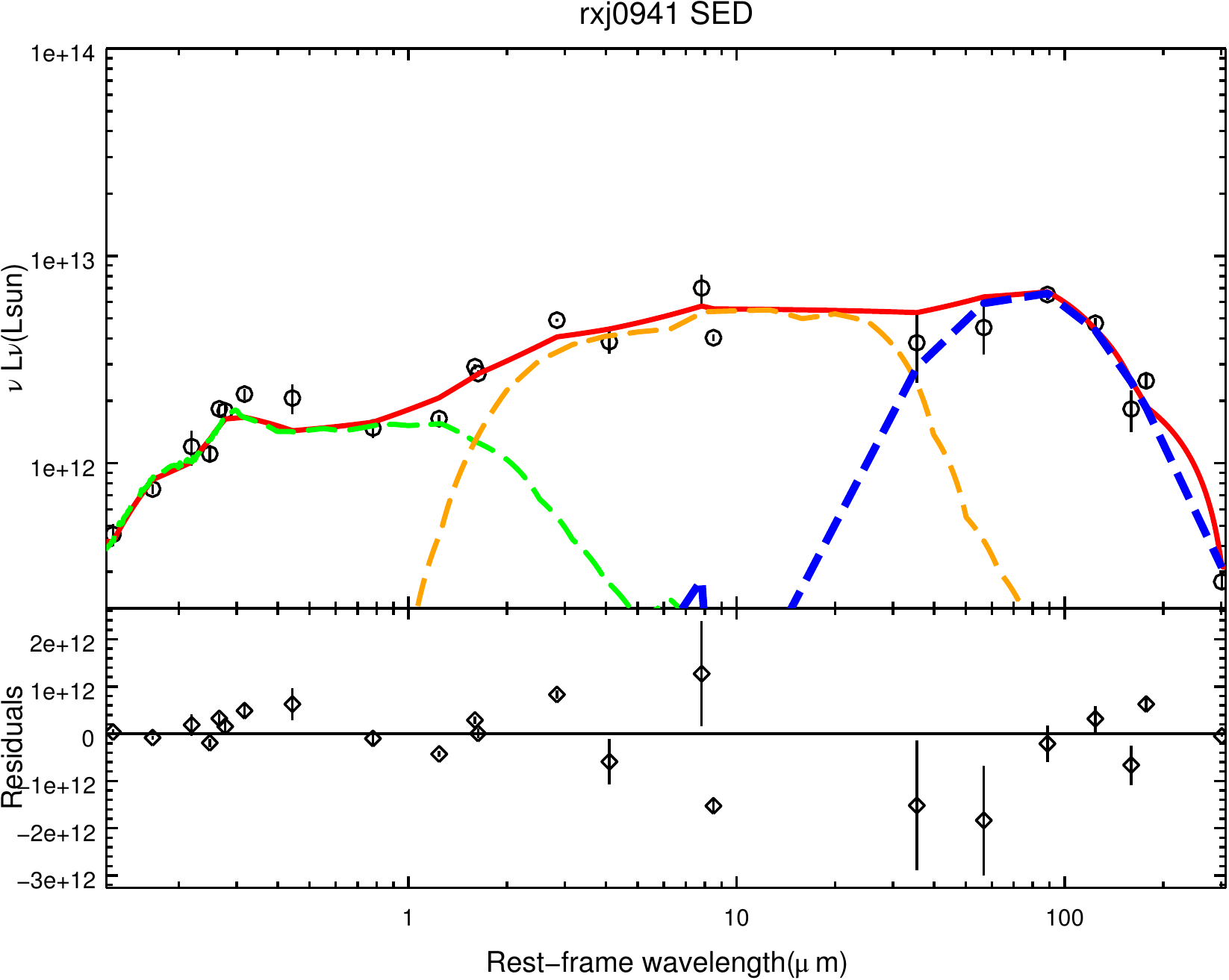}\\
   \includegraphics[width=0.45\textwidth]{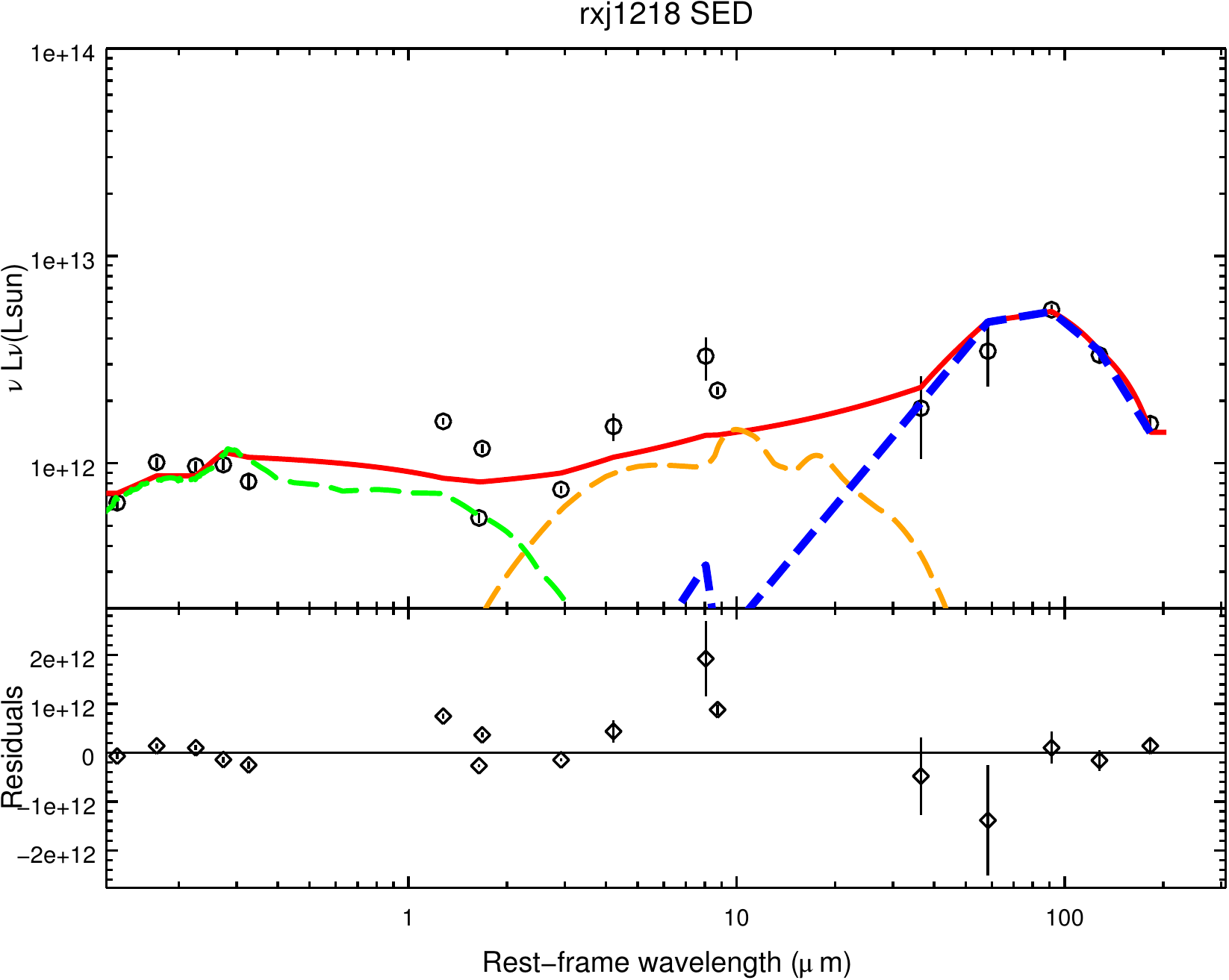}{\hspace {1cm}}
   \includegraphics[width=0.45\textwidth]{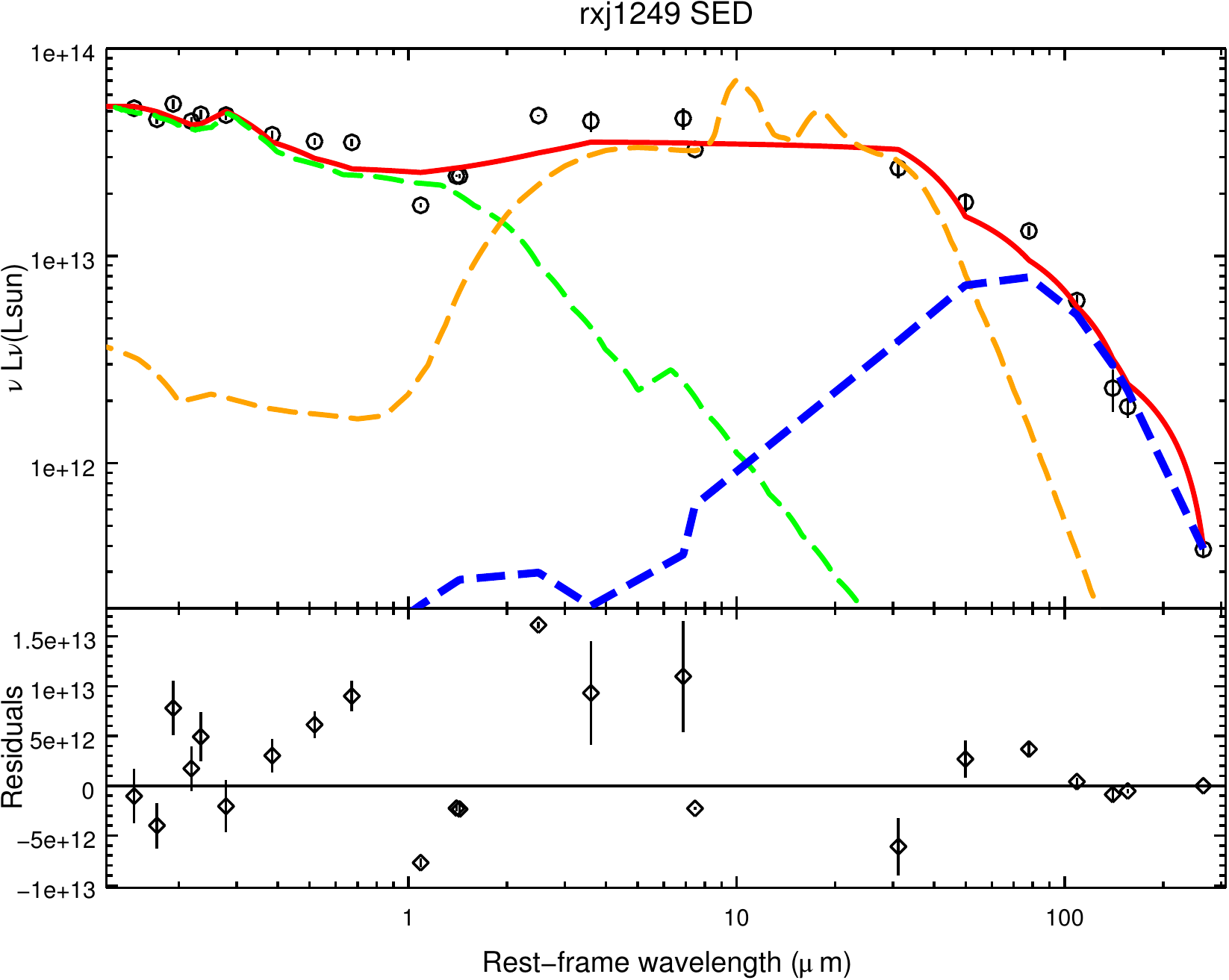}\\
   \includegraphics[width=0.45\textwidth]{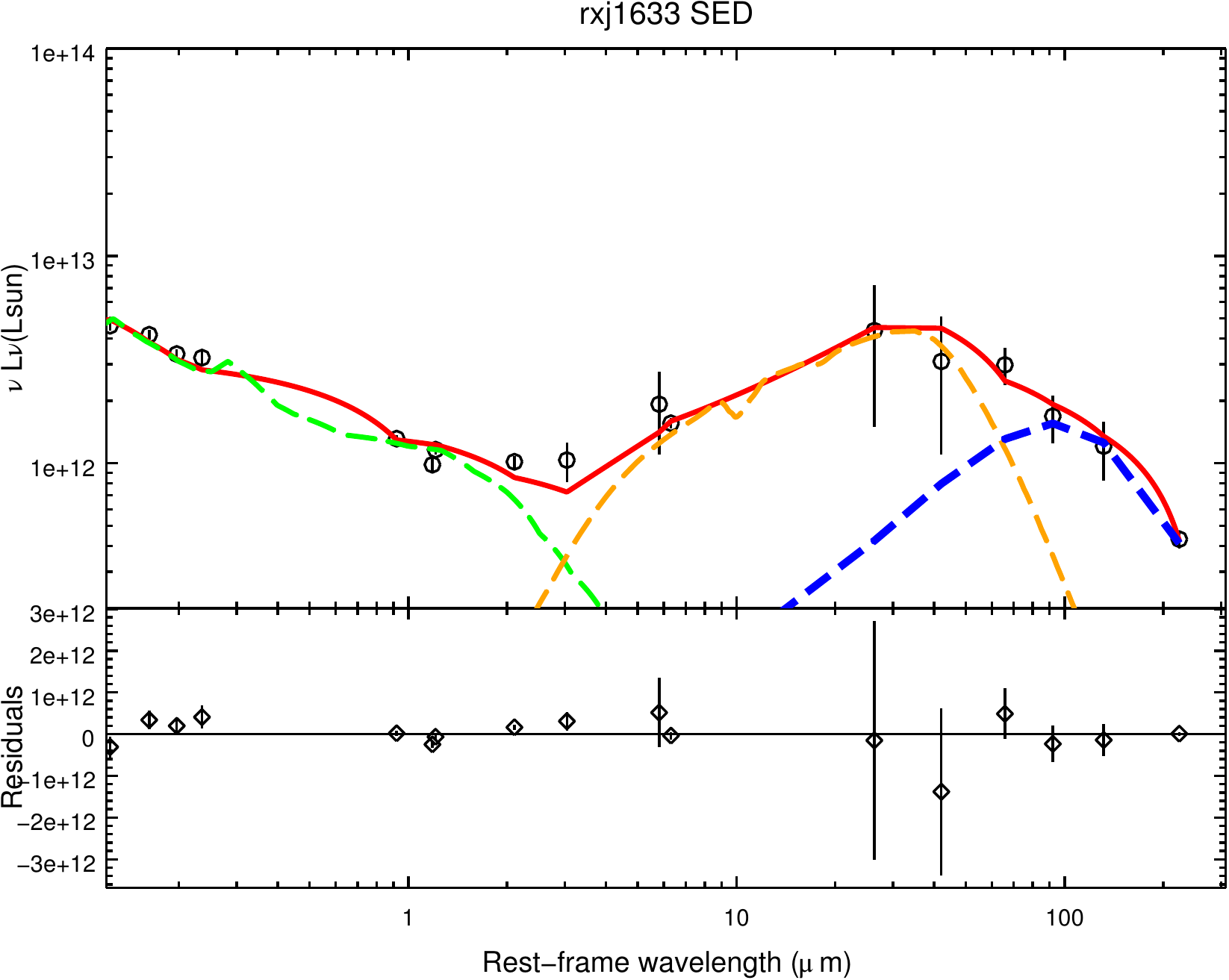}{\hspace {0.9cm}}
   \includegraphics[width=0.46\textwidth]{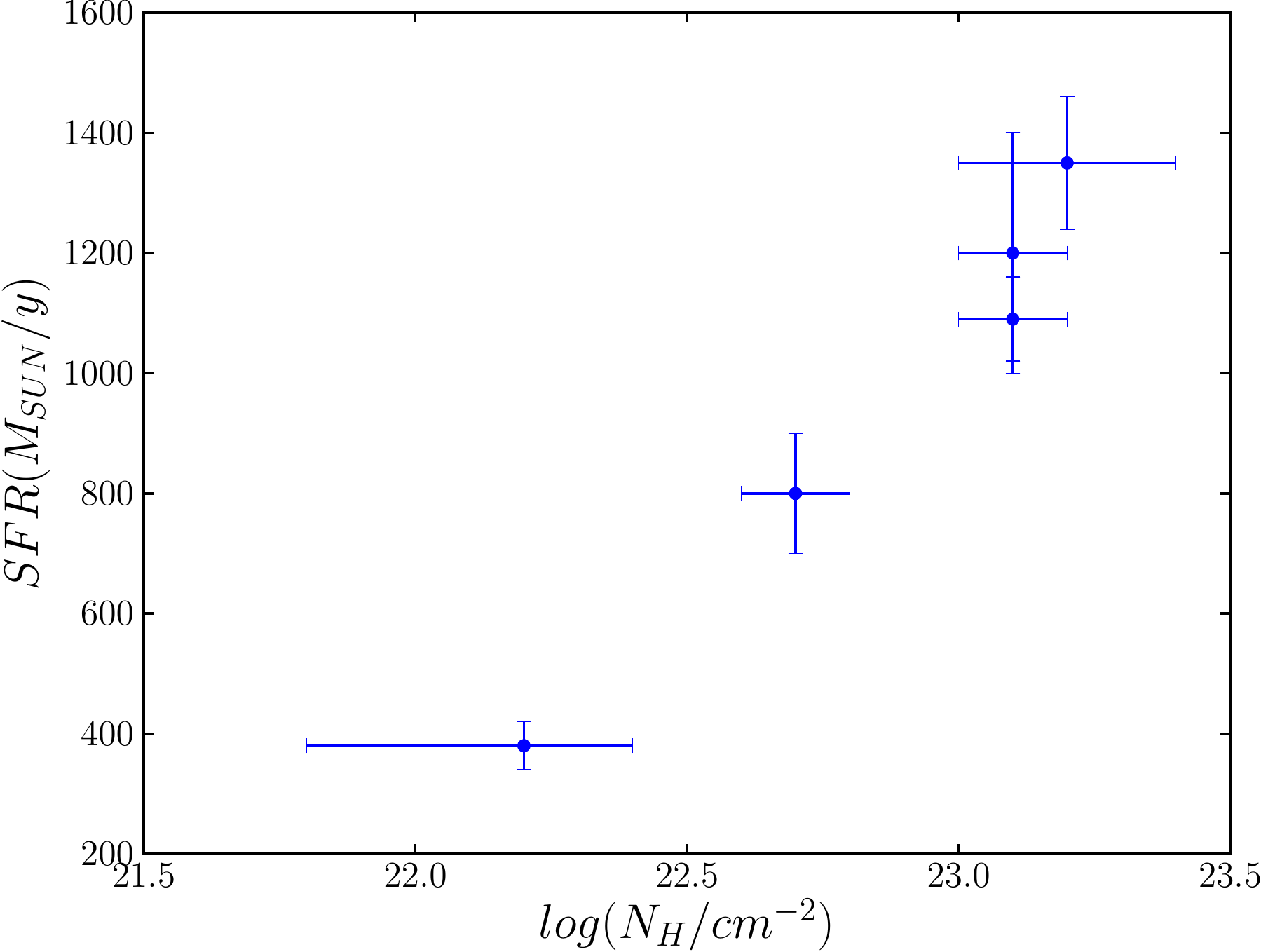}
   \caption{Each plot represents the combination of templates that
     best fits each SED. We show the total combined template (red solid 
     line) and each individual component: obscured direct accretion
     disk (green dashed line), reprocessed in the torus (orange dashed
     line) and star formation (blue dashed line). 
     For each source we also indicate below the torus component that provides the
     best-fit.  {\bf First row:} RX~J0057: Torus1 (left column). RX~J0941: Dusttor (right column). 
     {\bf Second row:} RX~J1218: Torus2 (left column). RX~J1249: Torus1 (right column).
     {\bf Third row:} RX~J1633: Torus3. (left column). Star formation rate versus N$_{H_{ion}}$ for our sample. (right column)}
  \label{figAllFits}
\end{figure*}

\begin{table*}
 \centering
 \caption{Average results and dispersions of fit parameters for each QSO, except for the X-ray AGN luminosity $L_{X,AGN}$ 
   which comes from \cite{page11}. $\chi^2$ reprents the minimum value for each best-fit family and $N$ represents the number of photometric points that we have for each object.}
 \begin{minipage}{180mm}
  \begin{tabular}{cccccccc}
  \hline
  \hline
  Object & $L_{X,AGN}$ &  $N_H$  & $L_{DISK}$ & $L_{TORUS}$  & $L_{FIR}$ & $L_{IR}$   & $\chi^2$ / N \\
  & ($10^{11} L_{\odot}$) &($10^{20}\, cm^{-2}$) & ($10^{11} L_{\odot}$) & ($10^{11} L_{\odot}$) & ($10^{11} L_{\odot}$) & ($10^{11} L_{\odot}$) & \\
 \hline
 RX~J0057 & 2.93 & $0.01 \pm 0.05$ & $125 \pm 2 $& $192 \pm 2$  & $49 \pm 6$ & $174 \pm 8$ &  159 / 20\\
 RX~J0941 & 0.93 & $0.77 \pm 0.11$ & $100 \pm 3$ & $148 \pm 9$ & $78 \pm 6$ & $179 \pm 13$  &  347 / 23 \\
 RX~J1218 & 2.33 & $0.39 \pm 0.08$ & $42 \pm 7$ & $25 \pm 3$ & $63 \pm 4$  & $90 \pm 11$ & 414 / 17 \\
 RX~J1249 & 3.69 & $0.14 \pm 0.03$& $1230 \pm 70$ & $1156 \pm 8$& $70 \pm 13$ & $782 \pm 19$ & 3109 / 23\\
 RX~J1633 & 5.85 & $0.003 \pm 0.002$ & $64.9 \pm 0.6$ & $74 \pm 5$  & $22\pm 2$ & $90 \pm 5$ & 29 / 17\\
\hline
\end{tabular}
    \label{LumResults}
\end{minipage}
\end{table*}

\section{Discussion}
In this Section we will piece together the clues obtained above about
the nature of our objects. We confirm the presence of strong FIR emission in these
objects (well above that expected from plausible AGN emission models) which 
we attribute to SF at the ULIRG/HLIRG level with SFR$\sim$1000 $M_{\odot}/$y.
Their associated greybody temperature values are close to those of
Submillimeter Galaxies (SMGs). They have dust masses around
$10^9$~M$_\odot$.

The black holes powering our QSOs are very massive at their epoch,
$\sim 10^{9}-10^{10} M_{\odot}$ (measured from broad emission lines in
their optical-UV spectra). We have calculated their mass-doubling
timescale $\tau$ and the time to reach the maximum BH mass observed
locally ($2\times 10^{10} M_{\odot}$), concluding that they can not grow much more. A further hint
in this direction comes from the high Eddington ratio of RX~J1633
which, according to \cite{kelly} should persist only for a very
brief period of time. RX~J1249 could become one of the most massive
objects known. We do not know the masses of their host galaxies, but their black hole
masses and their high SFR lead us to conclude that they are already
very massive or they would not have enough time to reach the local
bulge-to-black-hole-mass ratio. This is also in agreement with recent
models of AGN-host galaxy co-evolution, e.g. the recent recently proposed by \cite{lapi}. Our objects, with
$0.04<L_{FIR}/L_{BOL}<0.81$ would be in a stage when the FIR-luminous
phase has ended (Figure 15 in \cite{lapi}), in qualitative
agreement with our conclusion that their host galaxies are already
mostly formed.

\begin{table*}
 \centering
  \caption{Additional physical quantities derived from the average values and dispersions of the best-fit parameters and 
  Summary of timescales for QSOs and star formation, assuming
    constant mass accretion rates and SFR. $\tau$
    is the black hole mass-doubling timescale, and $\tau_{SB}$ is
    the time needed to reach the corresponding maximum host galaxy mass. $t$ is the ``look-back time''.}
 \begin{minipage}{180mm}
  \begin{tabular}{cccccccc}
  \hline
  \hline
  Object &  $L_{BOL}$ & CF &  $SFR$ &  $\log(M_{BH})$ &  $\tau$  & $\tau_{SB}$ & $t$\\
  & ($10^{11} L_{\odot}$) &  & ($M_{\odot}y^{-1}$)  & ($\log(M_{\odot})$) & $(Gy)$ & $(Gy)$ & $(Gy)$\\
 \hline
 RX~J0057 & $246.6\pm 1.0$ & $0.78 \pm 0.01$ & $800 \pm 100$ & 9.94$\pm$0.36 & $0.5_{-0.4}^{+0.3}$ & $10.7 \pm 1.3$ & 10.6 \\
 RX~J0941 & $169.3 \pm 1.2$ & $0.87 \pm 0.05$ & $1350 \pm 110$ & 9.77$\pm$0.40 & $0.5_{-0.4}^{+0.3}$ & $6.32 \pm 0.5$ & 10.0\\
 RX~J1218 & $78 \pm 4$ & $0.32 \pm 0.04$ & $1090 \pm 70$ & 9.28$\pm$0.45 & $0.4_{-0.3}^{+0.2}$ & $7.8 \pm 0.5$ &  9.9 \\
 RX~J1249 & $1890 \pm 30$ & $0.613 \pm 0.012$ & $1200 \pm 200$ & 9.99$\pm$0.45 & $0.08_{-0.07}^{+0.06}$ & $7.1 \pm 1.2$ & 10.6 \\
 RX~J1633 & $185.5 \pm 0.4$ & $0.40 \pm 0.03$ & $380 \pm 40$ &  8.73$\pm$0.36 & $0.04_{-0.04}^{+0.02}$ & $22 \pm 2$ & 11.3 \\
\hline
\end{tabular}
    \label{discu}
\end{minipage}
\begin{minipage}{180mm}
\end{minipage}
\end{table*}

Previous works have found some evidence of a correlation between the
SF of the host and the AGN obscuration in the X-rays (\cite{alexander}, \cite{bauer}, \cite{georgakakis},
\cite{rovilos07}). Later studies with deeper surveys both in the
X-ray and infrared ranges have failed to repoduce these results (e.g
\cite{rosario}, \cite{rovilos}).  We have therefore looked for a
correlation between N$_{H_{ion}}$ (from \cite{page11}) and SFR in
our sources (Fig. \ref{figAllFits} fith row right column), finding a tentative positive
correlation between these parameters. This is interesting, since it 
would imply a coupling of the ionized gas absorbing the X-rays at the scale 
of the accretion disk or the BLR with the gas forming stars in the host galaxy 
bulge, about three orders of magnitude farther away. At face value, this would be
compatible with a positive feedback scenario, \cite{king13} in which
the ionized outflowing gas would trigger star formation in the
interstellar medium of the host galaxy, with the highest column
density gas corresponding to stronger feedback.

%
%
\small  
%
\section*{Acknowledgments}   
A.K.A, F.J.C. and S.M. acknowledge financial
support from the Spanish Ministerio de Econom\'ia y Competitividad.
under project AYA2012-31447.  SM acknowledges Financial support from
the ARCHES project (7th Framework of the European Union, No. 313146).

%

%
\end{document}